\documentclass[a4paper,10pt]{article}
\usepackage{amsmath,amssymb}
\begin{document}
\def\nmonth{\ifcase\month\ \or January\or
   February\or March\or
April\or May\or June\or July\or August\or
   September\or October\or
November\else December\fi}
\def\nmonth{\ifcase\month\ \or January\or
   February\or March\or April\or May\or June\or July\or August\or
   September\or October\or November\else December\fi}
\def\rightheadline{\hfill\folio\hfill}
\def\leftheadline{\hfill\folio\hfill}
\newtheorem{theorem}{Theorem}[section]
\newtheorem{lemma}[theorem]{Lemma}
\newtheorem{remark}[theorem]{Remark}
\def\operatorname#1{{\rm#1\,}}
\def\text#1{{\hbox{#1}}}
\def\qedbox{\hbox{$\rlap{$\sqcap$}\sqcup$}}
\def\BB{{\mathcal{B}_1}}
\def\CC{{\mathcal{B}_2}}
\def\BX{{\mathcal{B}_{DR}}}
\def\BD{{\mathcal{B}_D}}
\def\BR{{\mathcal{B}_R}}
\def\B{{\mathcal{B}}}
\def\tr{{\operatorname{Tr}}}
\def\dvol{{\operatorname{dvol}}}
\newcommand{\reals}{\mathbf{R}}
\newcommand{\nats}{\mbox{${\rm I\!N }$}}

\def\gat{\gamma_a^T}
\def\la{\lambda}
\def\om{\omega}
\def\La{\Lambda}
\def\Th{\theta}
\def\chs{\chi^\star}
\def\Dirac{i\partial\!\!\!\!/}
\def\DDirac{iD\!\!\!\!/}
\def\dirac{i\partial\!\!\!\!/-A\!\!\!\!/}
\def\pip{\Pi_+}
\def\pim{\Pi_-}
\def\pipl{\Pi_+^\star}
\def\pimi{\Pi_-^\star}
\def\gf{\tilde{\gamma}}
\def\rand{\left|_{\partial M}\right. }
\def\ch{\cosh \Theta}
\def\sh{\sinh\Theta}
\def\nen{\Lambda \ch -2\om\sh}
\def\tr{\mbox{Tr}}
\def\lg{\lgroup}
\def\rg{\rgroup}
\def\R{\Re{\rm e}}
\def\I{\Im{\rm m}}
\def\ve{\varepsilon}
\def\wt{\widetilde}
\def\p{\partial}
\def\o{\overline}

\newcommand{\gc}{\gamma _M^0}
\newcommand{\gu}{\gamma _M^1}
\newcommand{\gd}{\gamma _M^2}
\newcommand{\gt}{\tilde{\gamma}_M}
\newcommand{\gce}{\gamma _0}
\newcommand{\gue}{\gamma _1}

\newcommand{\beq}{\begin{eqnarray}}
\newcommand{\eeq}{\end{eqnarray}}
\newcommand{\nn}{\nonumber}

\title{Symmetries and the cosmological constant puzzle}
\date{}
\author{
A.A. Andrianov$^{\,a,b}$,
F. Cannata$^{\,c,d}$,
P. Giacconi$^{\,c}$,\\
A.Yu. Kamenshchik$^{\,c,d,e}$,
R. Soldati$^{\,c,d}$}
\maketitle
\centerline{$^{a}$ {\it V.A. Fock Department of Theoretical Physics,}}
\centerline{\it 
Saint Petersburg State University, 198904, S.-Petersburg, Russia}
\centerline{$^{b}$ {\it Departament Estructura i Constituents de la Materia,}}
\centerline{\it  Universitat de Barcelona, 08028, Barcelona, Spain}
\centerline{$^{c}$ {\it Istituto Nazionale di Fisica Nucleare, 
                        Sezione di Bologna,}}
\centerline{\it 40126 Bologna, Italia}
\centerline{$^{d}$ {\it Dipartimento di Fisica,
Universit\'a
di Bologna}}
\centerline{$^{e}$ {\it L.D. Landau Institute for Theoretical Physics of the Russian
Academy of Sciences,}} 
\centerline{\it Kosygin str. 2, 119334 Moscow, Russia}

\begin{abstract}
We outline the evaluation of the cosmological constant in the framework of the 
standard field-theoretical treatment of vacuum energy and discuss the relation 
between the vacuum energy problem and the gauge-group 
spontaneous symmetry breaking. We suggest possible extensions  of the 't
Hooft-Nobbenhuis symmetry, in particular, 
its complexification till duality symmetry and discuss the compatible implementation on gravity.
We propose to use the discrete time-reflection transform to
formulate a framework 
in which one can eliminate the huge contributions of vacuum energy into 
the effective cosmological constant and suggest that the breaking of 
time--reflection symmetry could be responsible for 
a small observable value of this constant. 
\end{abstract}

\section{Introduction}
The so called cosmological constant problem has two quite different aspects, which are not always 
 clearly distinguished in the literature. One of these aspects is genuinely classical 
or even geometrical in its
origin. The corresponding question could be formulated as follows: 
the classical cosmological constant, which can be 
introduced into Einstein--Hilbert action and Einstein equations,
should be equal to zero and if it should, why?
In other words: does a symmetry exists which forces the vanishing of this constant?  
The second aspect is, instead, purely quantum-theoretical. Independently of 
the presence of the classical 
cosmological constant, the vacuum fluctuations of the quantum fields give the 
contribution to the energy-momentum tensor
which behaves as a cosmological constant, {\it i.e.} has the equation of state 
$p = -\rho$, where $p$ is pressure and 
$\rho$ is energy density. Naturally, this constribution is ultraviolet divergent. 
In the quantum field theory without 
gravity the problem is resolved by choosing the normal (Wick) ordering of creation 
and annihilation operators. This procedure 
is justified by fact that one measures the differences between energy levels, 
and not their absolute values. However, it is just the
absolute value of terms in the energy-momentum tensor which stays in the 
right-hand side of the Einstein equations and 
in the presence of gravity the Wick's normal--ordering loses its validity. 
A generally accepted procedure 
of the renormalization of vacuum energy does not exist, while the na\"\i ve cutoff 
imposed on the integration in the four-momentum space 
at the Planck scale gives huge values. How can one cope with them? 
One of possible approaches consists 
in the search of symmetries, which prohibit the existence of the cosmological 
constant and eliminate it on both the levels:
classical and quantum. It is well known that  supersymmetry suppresses 
divergences due to the compensating role of fermions 
and bosons. However, there are some difficulties at the application of the
supersymmetric models to both 
the cosmological constant problem and to the correct description of the
particle physics phenomenology, including the Higgs boson mass problem.  
Recently arguments were provided to force the vanishing of cosmological
constant even limiting oneself to the bosonic sector \cite{tHooft}. 
The formulation of this symmetry requires to give meaning to the space--time 
coordinates complexification (see also \cite{bonelli}, \cite{Erdem},
where a similar transformation was proposed but in six dimensions).
Here we would like to try to fold together the idea of complexification and 
the idea of compensation in a some new way. 

The structure of the paper is as follows: in Sec. 2 we outline the evaluation 
of the cosmological constant in the framework of the 
standard field-theoretical treatment of vacuum energy and discuss the relation 
between the vacuum energy problem and the gauge-group 
spontaneous symmetry breaking. In Sec. 3 we study the proposal \cite{tHooft} 
and its possible extensions, starting from the formalism 
developed in Sec. 2, in particular, its complexification allows to introduce a
suitable duality symmetry. Its effect on gravity 
is also discussed and possible ways to implement a correct gravity interaction 
are outlined.
Finally, in Sec. 4 we propose to use  the discrete time reflection to
formulate a framework in which one can eliminate 
the huge contributions of vacuum energy into 
the effective cosmological constant and, moreover, we suggest that  
the breaking of the T symmetry could be responsible for 
a small observable value of this constant.  

\section{Renormalization of vacuum energy density and spontaneous symmetry breaking}

Recall that
 summing the zero--point energies of all the normal
modes of some field component of mass $m$ up to a wave number cutoff
$K\gg m$ yields a vacuum energy density
\beq
\langle\,\rho (m)\,\rangle =\frac12\,\int_0^K\frac{4\pi k^2dk}{ (2\pi)^3}\,
\sqrt{k^2+m^2} =\ \frac{K^4}{16\pi^2}+\frac{K^2m^2}{16\pi^2}-\frac{m^4}{32\pi^2}
\left[\,\ln\frac{K}{m}-\frac14+\ln2+O\left(\frac{m}{K}\right)^2\,\right]\ . \label{zeropoint}
\eeq
If we trust in general relativity up to the Planck scale $M_p$, we might take
$K\simeq M_p = (8\pi G_{\cal N})^{-1/2}\,,$ which would give
\beq
\langle\,\rho\,\rangle\approx 2^{-10}\,\pi^{-4}\,G_{\cal N}^{-2}=2\times 10^{71}\ {\rm GeV}^4\ .
\eeq
But it is known that the observable value of the effective cosmological constant is  
less than about $10^{-47}\ {\rm GeV}^4\,,$ and, sometimes one writes that a huge fine-tuning seems to be at work.
However, it is necessary to be careful with such statements \footnote{We are grateful to A.A. Starobinsky, who has attracted our attention to this problem.} because naively cutoffed expression (\ref{zeropoint}) does not correspond to the cosmological constant. Within the same regularization one can calculate the vacuum pressure,
\beq
\langle\, p(m)\,\rangle \equiv \langle\,\frac13 T^j_{j}\,\rangle =
  \frac{1}{3} (1 - m \partial_m) \langle\,\rho (m)\,\rangle
\ . \label{zeropointflux}
\eeq
When comparing \eqref{zeropoint} with \eqref{zeropointflux} one finds that the quartic divergences behave like radiation $p = \rho/3$, the quadratic ones as a perfect fluid with the equation of state $p = -\rho/3$ and the logarithmic divergences reproduce the cosmological constant equation of state $p = -\rho$. Evidently first two components are not entirely Lorentz-invariant but are determined in the rest frame of the Universe.  
On the other hand, as was pointed out by Zeldovich \cite{Zeld} the vacuum expectation of the energy-momentum tensor should be Lorentz-invariant and that means that it should be proportional to the metric tensor. That implies that the pressure is equal to the energy density taken with the opposite sign, or in other words, it means that that vacuum energy-momentum tensor must behave as a cosmological constant. The Lorentz- invariant part can unambiguosly separated by averaging  different components of the energy-momentum tensor over Lorentz transformations and it does not include any radiation background thereby starting from quadratic divergences only 
 (see similar arguments in \cite{Akhmedov}). Still the vacuum energy remains huge as compared
to the energy density related to observable cosmological constant.

Meantime, in \cite{Zeld} it was shown that requiring the elimination of all the divergences due to some general renormalization 
procedure, equivalent to introducing a spectral function of some kind, one automatically deduces that the finite part of 
the energy-momentum tensor have a Lorentz-invariant form. Considering the spectral function not as a renormalization tool, but
as giving a real particle specrtrum, one can have a general restrictions on the particle physics models, providing the 
cancellation of the ultraviolet divergences in the energy-momentum tensor not only on the Minkowski, but also on the de Sitter 
background \cite{Venturi}. 
 Here we discuss how various forms of regularization can be used to control the UV divergences.

We can  try to regularize the zero--point energy density in terms 
of the dimensional regularization \cite{birrel}: namely,
\beq
\langle\,\rho\,\rangle = \frac12\,(2\pi)^{1-n}\int d^{\,n-1}k\
(k^2+m^2)^\frac12 = \frac12\,(2\pi)^{1-n}\,\frac{2\pi^{(n-1)/2}}{\Gamma[\,(n-1)/2\,]}
\int_0^\infty dk\ k^{n-2}(k^2+m^2)^\frac12\ .
\eeq
Unfortunately there is no strip in the complex $n$--plane
in which the above integral is well defined, so that dimensional
regularization is not appropriate in order to give a meaning
to the zero--point energy.

\medskip
Alternatively we could also define the zero--point energy density in the
path--integral formalism \cite{birrel}, which turns out to be quite 
convenient in view of its
generalization to the curved space. 
Consider the classical action
\beq
S[\,\phi\,]\ =\ \frac12\int d^{\,4} x\ [\,\partial_\mu\phi\,\partial^{\,\mu}\phi
- m^2\,\phi^2\,]\ ,
\eeq
and define the kinetic invertible operator,
\beq
{\cal K}_{x}\ :=\ (\Box_x+m^2-i\ve) ,\label{prop1}
\eeq
and its Feynman propagator ,
\beq
G_F(x-y) \equiv -\ {\cal K}_{xy}^{\,-1} = \int\frac{d^{\,4}k}{(2\pi)^4}\ (k^2-m^2+i\ve)^{-1}
\exp\{ik\cdot(x-y)\}\ .\label{prop2}
\eeq
Then we find the generating functional 
\beq
Z[\,J\,]&:=&\int \mathfrak{D}[\,\phi\,]\ \exp\left\{
i\,S\,[\,\phi\,]+i\int d^{\,4} x\ J(x)\,\phi(x)\right\}\\ 
&:=& Z[\,0\,]\exp\left[\,-\,\frac{i}{2}\int d^{\,4} x\int d^{\,4} y\
J(x)\,G_F(x-y)\,J(y)\,\right]\ .\nn
\eeq
Now, in order to end up with a dimensionless 
generating functional, we can formally write
\beq
Z\,[\,0\,]&=&{\mathcal N}\,(\det||\,\mu_0^{-2}\,{\cal K}\,||)^{\,-1/2}
 = {\mathcal N}\,\exp\,[\,1/2\,\tr\,\ln\,(\mu_0^{\,2}\,{\cal K}^{\,-1})\,]\ \ ,
\label{vacuum}
\eeq
where ${\mathcal N}$ is an irrelevant numerical normalization constant
that we shall omit in the sequel, whereas $\mu_0$ is an arbitrary wave number or momentum scale.

A first possibility is to understand the formal relationships (\ref{vacuum})
in terms of the $\zeta$--function regularization \cite{hawking} that yields
\beq
\ln\,Z[\,0\,]&=&\frac12\,L^4\int\frac{d^{\,4}k}{(2\pi)^4}\ \frac{d}{ds}
\left[\,\mu_0^{-2}\,(-k^2+m^2)\,\right]_{s=0}^{\,-s}\ .
\eeq
After changing the integration variable $k^0=ik_4$ we find
\beq
\ln\,Z[\,0\,]\ =\ i\ \frac{m^4 L^4}{32\pi^2}\ \frac{d}{ds}\left\lgroup
\frac{\Gamma(s-2)}{\Gamma(s)}\,\left(m/\mu_0\right)^{\,-2s}
\right\rgroup_{s=0}\ ,
\eeq
so that we can eventually write
\beq
 Z[\,0\,]={\rm e}^{iW}\ =\
\exp\left\{-i\,L^4\,\langle\,\rho\,\rangle_{\rm eff}\right\}\ ,\qquad
 \langle\,\rho\,\rangle_{\rm eff} = \frac{m^4}{64\pi^2}
\left[\,\ln\,(m^2/\mu_0^{\,2})-3/2\,\right]\ .
\eeq
We see that the $\zeta$--function regularization drives to
a result for the zero--point vacuum energy density
which turns out to be {\sl IR logarithmically divergent and positive} 
when the infrared cutoff $\mu_0\varpropto L^{-1}$ is removed. It seems that
the $\zeta$--function regularization is not adequate in treating the cosmological
constant problem as it re-directs the problem from UV to IR region.

\medskip  
Turning back to eq.~(\ref{vacuum}), a second possibility is to use the
ultraviolet cutoff regularization of the large wave number field modes: 
namely,
\beq
\tr\Big(\, \vartheta(Q-{\cal K})\,\ln({\mathcal K}/\mu^2_0)\Big) =\ 
\frac{i\,L^4}{16\pi^2}\int_0^{\,Q-m^2}q\,\ln\,[\,(q+m^2)/\mu^2_0\,]\,dq
\eeq  
where $k^0=ik_4\,,\ k_E^{\,2}={\bf k}^2+k_4^2\,,\ Q\sim M_p^{\,2}\,,$  
so that we eventually obtain
\beq
\langle\,\rho\,\rangle_{\rm eff}\ =\ \frac{1}{128\pi^2}
\Big[\,Q^2 (2\ln\,[\,Q^2/\mu^2_0\,] - 1) -4Q\,m^2 (\ln\,[\,Q^2/\mu^2_0\,] - 1)+2m^4\ln\,(m^2/\mu^2_0)-3m^4\,\Big] 
\label{cutenergy}
\eeq
in a satisfactory agreement with eq.~(\ref{zeropoint}) up to a redefinition of the large wave
number cutoff.
Then one could  fit the
Lorentz invariant part of (1)  to (13)
choosing
$\ln Q/\mu^2_0 = 1/2$ as
for this choice the contribution $\sim \Lambda^4$ also vanishes in (13) .

\medskip
To be consistent with the standard model of particle physics we have to take care 
of spontaneous symmetry breaking in the field theory.
In this framework the charged scalar field potential takes the form 
(with $\mu^2>0,\, \lambda>0$) 
\beq
V(\phi)=V_0-\mu^2\phi^\dagger\phi+\lambda(\phi^\dagger\phi)^2\ .
\label{pot}
\eeq
The classical minimum of this potential occurs at the constant field 
values $\phi^\dagger\phi=\mu^2/2\lambda$ so that it is convenient to
parameterize the scalar field $\phi$ by writing
\beq
\phi(x)=U(x)\,\frac{1}{\sqrt2}\,\left\lgroup\begin{array}{cc}
  0 \\
  v+\sigma(x)\\
\end{array}\right\rgroup\ .
\eeq
We can now make a gauge transformation in order to eliminate $U(x)$
from the lagrangian -- unitary gauge -- in such a way that 
\beq
V_{\rm min}=V_0-\frac{\mu^4}{4\lambda}\ .
\eeq
According to the review \cite{steven} it is apparently suggested  
that the classical potential
should vanish at $\phi=0,$ which would give $V_0=0,$ so that some classical
negative contribution to the zero-point energy density would be there.
In the electroweak theory, if we assume an Higgs boson mass
$m_H=\mu\sqrt2\simeq 150$ GeV, this would give 
$\rho_0\simeq -(150\ {\rm GeV})^{\,4}/16\lambda\,,$
which even for $\lambda$ as small as $\alpha^2$ would yield 
$|\,\rho_0\,|\simeq 10^{\,12}\ {\rm GeV}^{\,4}$, larger than the observed value by a factor
$10^{\,59}$. Of course we know of no reason why $V_0$ or $\Lambda$ must vanish, and
it is quite possible that $V_0$ or $\Lambda$ cancels the term $-\mu^{\,4}/\lambda$
(and higher order corrections), but this example neatly shows how un--natural
is to get a reasonably small effective cosmological constant.
  
In general, if we turn to the shifted field $\sigma(x)$ in the unitary gauge,
we obtain the Lagrange density for the shifted field
\beq
{\cal L}[\,\sigma\,]=\frac12\,\partial_\mu\sigma\,\partial^{\,\mu}\sigma
-\frac12\,(2\mu^2)\,\sigma^2\mp\mu\sqrt\lambda\,\sigma^3-\frac{\lambda}{4}\,\sigma^4
-V_0+\frac{\mu^{\,4}}{4\lambda}\ .
\label{shiftedlag}
\eeq
Accordingly, the zero--point energy density in the symmetry broken phase
appears to be
\beq
\langle\,\rho\,\rangle = 
\langle\,\rho\,\rangle_{\rm div}
+V_0-\frac{\mu^4}{4\lambda}\ .
\eeq

\medskip
It is clear that we can easily remove 
the divergent part of the zero--point energy density of the scalar field $\sigma(x)$
after the introduction of a real {\sl mirror free real scalar field} $\varphi(x)$ 
with a classical Lagrange density related to that one of eq.~(\ref{shiftedlag}) 
\beq
{\cal L}\,[\,\varphi\,]=\frac12\,\partial_\mu\varphi\,\partial^{\,\mu}\varphi
-\frac12\,(2\mu^2)\,\varphi^2\ ,
\eeq
together with a {\sl ghost pair of scalar fields}. These mirror and ghost fields are supposed
not to interact directly to the SM fields. At the classical level
the ghost fields  are described by anticommuting, real Grassmann algebra valued, 
field functions $\eta(x)=\eta^\dagger(x),\,{\bar\eta}(x)=\bar\eta^{\,\dagger}$ with Lagrange density
\beq
{\cal L}_{\rm GP}\ =\ -\,i\,\p_\mu\bar{\eta}\,\p^{\,\mu}\eta
+2i\mu^2\,\bar{\eta}\,\eta\ .
\eeq
We have
\beq
\Pi_{\,\eta}\ =\ +\,i\,\p_0\bar\eta\ ,\qquad \Pi_{\,\overline{\eta}}= -\,i\,\p_0\eta\ ,
\eeq
so that correspondingly
\beq
H_{\rm GP}&=&\int d{\bf x}\left[\,\dot\eta\,\Pi_{\,\eta}
+\dot{\bar\eta}\,\Pi_{\,\overline{\eta}}
-{\cal L}_{\rm GP}\,\right]
= -\,i\,\int d{\bf x}\left[\,\p_0\bar\eta\,\p_0\eta + \nabla\bar\eta\cdot\nabla\eta
+2\mu^2\,\bar\eta\,\eta\,\right]\ . 
\eeq
Now one can quantize the ghost pair with the help of canonical anti--commutation relations .
As a consequence, after Fourier decomposition of ghost fields we eventually obtain
\beq
H_{\rm GP}&=& 
i\int d{\bf p}\ p_0\,
[\,\bar\eta^{\,\dagger}({\bf p})\eta\,({\bf p})-
\eta^{\,\dagger}({\bf p})\bar\eta\,({\bf p}) + i\,(L/2\pi)^3\,]\ ,
\eeq
so that we get the ghost pair negative contribution to the zero point
energy density in the large wave number cutoff regularization with the 
Planck mass
\beq
\langle\,\rho\,\rangle_{\,\rm GP} = -\,(2\pi)^{-3}
\int_0^{M_P}{4\pi p^2dp}\ (p^2+2\mu^2)^\frac12\ .
\eeq 
Thus the above mirror--symmetry, which does not mix the standard model
multiplets, would 
admittedly resolve the
cosmological constant problem  
in the scalar Higgs sector only. It can be extended onto the entire
field content of the standard model, at the expense of introducing more ghost fields with wrong
spin-statistics relation. 
Evidently, gravity will mix the standard model fields with their related
mirror--replic\ae\ and ghost--pair, leading eventually to the breaking of the
spin-statistics relation  and even unitarity in the standard model world. 
A rather sophisticated proposal will be formulated in the next section to skip
that nasty mixing and unitarity loss.
\section{
't Hooft-Nobbenhuis symmetry and cosmological constant}
In this section we 
explore the symmetry against the change of the full metric sign by
continuation of real space-time variables 
to complex values -- in the original proposal \cite{tHooft} to imaginary ones -- 
first in the flat Minkowski space-time,
\beq	
\eta_{\mu\nu} = \mbox{diag}\,\parallel +,-,-,-\,\parallel\,;\qquad x^{\,\mu} \mapsto -i
y^{\,\mu}\,,\qquad y^{\,\mu}= y^{\,\mu\,*}\,;\qquad 
\partial_\mu \mapsto i \partial_\mu\,;\qquad k_\mu \mapsto i k_\mu\ .
\eeq
Moreover, for a real scalar field $\phi(x)$ we shall set :
\beq 
\phi (x)=\phi (-i y)\quad \mapsto\quad \tilde\phi (y) =  \tilde\phi^*
(y)\,;\qquad \int d^4 x \mapsto \int d^4 y\ ;
\eeq
We stress that $\tilde\phi (y)\not= \phi (-i y)$ since $\tilde\phi(y)$
is evidently real, whereas $\phi (-iy)$ is in general complex. 
Therefore, the 't Hooft-Nobbenhuis transformation is not merely an analytic
continuation, as it involves an essential change of the functional base--space. 
It is of course analogous to what we do when we make the transition to the
Euclidean formulation, {\it e.g.} $\phi(t)\mapsto\phi(\tau = - it)$, in which
we perform a simultaneous mapping of one functional space -- a subspace of the complex function space, 
to another one spanned by real functions $\tilde\phi(\tau)$ of the Euclidean--time coordinate $\tau$.

In so doing, one finally comes to a theory of scalar tachyon -- this is the
reason why t' Hooft and Nobbenhuis actually neglect masses -- namely,
\beq
{\cal L}_x = \frac12\,\left(\partial_\mu\phi\,\partial^\mu\phi - m^2\phi^2\right) 
-\frac{\lambda}{4}\,\phi^4 \ \longmapsto \ -\,{\cal L}_y\,;\qquad 
{\cal L}_y \equiv \frac12\,\left(\partial_\mu\phi\,\partial^\mu\phi 
+ m^2 \phi^2\right) +\frac{\lambda}{4}\,\phi^4\  .
\eeq
with a repulsive quartic self-interaction, the issue of the vacuum stability
against small time-dependent perturbations keeping admittedly open.

We also remark that if the space-time is not flat, then the continuation
towards complex coordinates makes the metric also complex. However,
since the background metric is the solution of Einstein equation,  it can not
be na\"ively defined by analytic continuation. 
Rather,
one has to perform the corresponding mapping of the metric functional base--space
and then solve the Einstein equations 
with a transformed new energy--momentum tensor, because the matter fields are
in turn suitably mapped.
As a result, one might expect the very same background metric, if at least the classical
matter distribution remains unchanged, what is far from being obvious 
for massive interacting matter fields.
\subsubsection*{Our complementary proposal: extended duality symmetry}
Once that the 't Hooft--Nobbenhuis proposal necessarily involves the mapping of the functional
base--space, 
one could naturally include into this mapping also the analytic continuation
of the field variables. To this concern, one could treat 
all the field amplitudes and coordinates, but the metric, on the same footing: 
{\it e.g.} for real scalar fields,
\beq
\phi (x)=\phi(-i y)\ \mapsto\ i\varphi (y)\,; \qquad 
\varphi (y) = \varphi^* (y)\,;\quad {\cal L}_y (\varphi) = 
\frac12\,\left(\partial_\mu\varphi\,\partial^\mu\varphi + 
m^2 \varphi^2\right) -\frac{\lambda}{4}\,\varphi^4\ .
\eeq
This transformation does realize a link between two scalar field theories 
without and with spontaneous symmetry breaking . The latter one
has its classical  minima at $\left\langle \varphi \right\rangle = \pm\,{m}\lambda^{-1/2}\,.$ 
After the field amplitude shift
$\varphi \mapsto \left\langle \varphi \right\rangle + \varphi\,,$ the effective lagrangian reads
\beq
{\cal L}_y (\varphi) = \frac12\,\left(\partial_\mu\varphi\,\partial^\mu\varphi
- 2 m^2 \varphi^2\right) - \frac{\lambda}{4}\,\varphi^4 \mp m \sqrt{\lambda}\,
\varphi^3\ .
\eeq
Moreover, for vector gauge fields we shall write
\beq
A_\mu(x)\ \mapsto\ i\,{V}_\mu (y)\,;\quad D^{\,x}_\mu = \partial^{\,x}_\mu + i A_\mu(x)\
\mapsto\ i\,{D}^{\,y}_\mu 
=i\,[\,\partial^{\,y}_\mu + i\,{V}_\mu(y)\,]\,;\quad F_{\mu\nu}(x) \mapsto -\,{G}_{\mu\nu}(y)
\eeq
to preserve covariant derivatives. The above transformations just leave both 
the Maxwell's lagrangian and the action invariant.
Finally, for massless fermions with Yukawa coupling $g$
\beq
\psi(x)\ \mapsto\ \psi(-iy)\ \mapsto\ \Psi (y)\ ;\qquad \bar\psi(x)\ \mapsto\
-\,i\bar\psi(-iy)\ \mapsto\ \bar\Psi (y) 
= \Psi^\dag\,\gamma^0
\eeq
\beq  
{\cal L}_x = \bar\psi\left( i\not\!\partial - 
{\not\!\! A} - g\,\phi\right) \psi\  \longmapsto \  {\cal L}_y = 
\bar\Psi\left( i\not\!\!\partial - {\not\! V} - g\,\varphi\right) \Psi
\eeq
so that the spontaneous symmetry breaking mechanism allows to generate
the correct  fermion masses. We remark, had one included the bare
fermion  masses, then they would become imaginary with the above rules. Therefore the
transformed  fermions would
be unstable, having the imaginary bare part and the real part generated by
spontaneous symmetry breaking.
\subsubsection*{Vacuum energy under the extended duality symmetry}
Suppose that the vacuum energy is compensated to {\it zero} in the
original theory. 
This compensation can be described in an effective theory style by introducing
a number 
of {\it shadow}  fields -- in analogy with the  Pauli-Villars regularization scheme --
with the same coupling constants but different masses. 
For instance, a real scalar field $\phi$ is supplemented
by $N$ shadow fields $\varphi_j\,,\ j=1,2,\ldots,N$ with masses $M_j$, the same quartic coupling
constant 
and the same effective lagrangian but with a positive or negative  
weight of their contribution into the effective action $\Gamma$
\beq
\Gamma = 
\Gamma (\phi,m,\lambda) - \sum\limits_{j=1}^N (-1)^{\,p_j}\Gamma(\varphi_j, M_j,\lambda) ;
\qquad p_j = 1,2; \label{shadow}
\eeq
where we keep the same notation $\phi,\varphi_j$ to indicate the classical
mean field variables in the Legendre functional transform.
We stress that they are combined, at the level of the effective action, in order 
to exactly compensate the vacuum energy contributions. All of them are bosons, although 
some of them behave as ghosts in the leading {\it quasi}--classical approximation.

Within the framework of quantum field theory, 
one can interpret the part of this set with negative sign (even $p_j $)
as originating from the {\it evolution backward in time} with anticausal prescription for propagators,
{\it {\it i.e.}} with replacing $+i\ve$ into $-i\ve$ in \eqref{prop1},\eqref{prop2} 
 -- see an example in \cite{Sundrum}.

Concerning the zero--point energy, one normally has three types of leading divergences -- 
see eq.~(\ref{cutenergy}) 
$$
\sim Q^{\,2}\qquad \sim Q\,m^2\qquad \sim m^4 \log\,(Q/m^2)
$$
and with the help of a number of shadow fields one exactly cancels the divergences if
\beq
\sum\limits_{j=1}^N (-1)^{\,p_j}  = 1,\qquad   
\sum\limits_{j=0}^N (-1)^{\,p_j} M^2_j = 0,\qquad p_0 = 1\qquad M_0 \equiv m,\qquad  
\sum\limits_{j=0}^N (-1)^{\,p_j} M^4_j= 0\ . \label{cancel}
\eeq
We assume of course the preliminary mass and coupling constant renormalization.
One can also assume that the shadow world consists of sufficiently heavy particles in order to reduce 
their influence as much as possible on the physics accessible in the standard model real  world. 
Then the minimal number of such fields is equal to five. 
The first sum rule can be interpreted as a ``conservation law" of a number of matter sub-worlds evolving forward and backward in time in a certain accordance with no-time origin of our universe
\cite{rep-grav}.

On the one hand, once the light shadow fields has been accepted, 
one can restrict himself to solely one
species with negative sign of its effective action -- compare with \cite{Sundrum}.

On the other hand, the cancellation of quartic and quadratic divergences has to
be resorted to Planck's scale physics, where the very notion of low energy fields 
with their Lagrangians of canonical dimension four is admittedly questionable. 
A self--consistent treatment at low energies must deal then
with light mass scales and relatively light shadow fields (as compared to the Planck mass) 
and therefore only with the last relation, which involves the fourth powers of the shadow masses. 

After the duality transformation $\phi_j \mapsto \varphi_j$ and the resolution of
spontaneous symmetry breaking by 
shifting each field in $\left\langle \varphi_j \right\rangle =
\pm\,{M_j}/{\sqrt{\lambda}}\,,$ 
one finds the classical vacuum energy density
\beq
\langle\,\rho\,\rangle_{cl} = \sum\limits_{j=0}^N (-1)^{\,p_j}\left[\,-\,\frac12\,M^2_j
\left\langle \varphi_j 
\right\rangle^2 +\frac{\lambda}{4}\left\langle \varphi_j
\right\rangle^4\,\right] 
= -\,\frac{1}{4\lambda}\sum\limits_{j=0}^N  (-1)^{\,p_j} M^4_j = 0\ .
\eeq
Thus, quite remarkably, after the extended duality transformation 
the scalar field vacuum energy density remains equal to zero, 
whereas the masses are generated, both for fermions and for gauge bosons, thanks to  
the Higgs mechanism. Certainly all the standard model fields must be replicated in the shadow
sectors, if one provides the zero cosmological constant. If those shadow fields do not
interact with each other, it cannot be conceivably embedded into a minimal supersymmetry. 
On the contrary, if one starts from a minimal exact supersymmetry, 
this dressing by shadow fields and the 
subsequent extended duality transformation might lead to a 
spontaneous symmetry breaking for supersymmetry with zero cosmological constant in the outcome.
\subsubsection*{Hints for gravity}
Suppose that shadow fields interact with our world only through gravity 
and therefore they belong to the dark side of the universe.
Then we could exploit the shadow fields as a part of the matter in the universe and 
not merely just like regularizing fields.
Since after the extended duality transformation we change the sign of derivatives 
but not of the metric, we have
\beq
R_{\mu\nu}(x)\quad\longmapsto\quad -\,R_{\mu\nu}(y)\eeq 
\beq S_{g} = -\,\frac{1}{16\pi G_{\cal N}}
\int d^4 x\ [\,R(x) - 2\Lambda)\,]\quad\longmapsto\quad  
\frac{1}{16\pi G_{\cal N}} \int d^4 y\ [\,R(y) + 2\Lambda\,]\ .
\eeq
There is no invariance, as the sign of cosmological constant  is unchanged albeit gravity becomes 
anti--gravity.
The possible solutions are:
\begin{enumerate}
\item anti--gravity in the symmetric phase $G_{\cal N} < 0$ (repulsion supports this phase) is
replaced by true gravity in the spontaneous symmetry broken phase, although  
one has to check classical solutions ;
\item gravity is induced solely by matter and therefore the overall sign of the 
gravitational action remains the same
under the extended symmetry transformation and the spontaneous symmetry breaking,
albeit the compensation mechanism, if it is exact, does select out a vanishing coefficient 
in front of the scalar curvature ;
\item there is a coupling to scalar fields: namely,
\end{enumerate}
\beq
S_{g} &=& - \int d^4 x\,\Big(A + \sum\limits_{j=0}^N B_j\,\phi_j^2(x)\Big)\,R(x)\
\longmapsto\  
\int d^4 y\,\Big(-\,A + \sum\limits_{j=0}^N B_j\,\varphi_j^2(y)\Big)\,R(y)\nonumber\\
&=& \Big(A - \frac{1}{\lambda}\sum\limits_{j=0}^N B_j M^2_j\Big)\,\int d^4 y\,R(y)\ +\ \ldots\
=\ \frac{1}{16\pi G_{\cal N}}\,\int d^4 y\,R(y)\ +\ \ldots
\eeq
so that
\beq
A = 
\frac{1}{16\pi G_{\cal N}}\ ;\qquad  
\sum\limits_{j=0}^N B_j M^2_j =  
\frac{\lambda}{8\pi G_{\cal N}}\ .
\eeq
If gravity is not principally induced by matter fields, then in the latter case there is
no prescribed relation between the individual gravitational scalar couplings $B_j\,.$
In such a circumstance, one could adjust them to support essential invariance under the 
extended duality transformation and a tiny cosmological constant might be generated 
via vacuum polarization.

\section{Cosmological constant, time arrow and T violation}
First of all, let us remark that according to recent observational results 
such as the discovery of the cosmic acceleration 
\cite{accel} it is reasonable to think that the real value of the cosmological 
constant is not strictly zero. Indeed, the so called 
$\Lambda$CDM cosmological model based on the presence of the cosmological constant 
has acquired the status of the standard 
cosmological model.  Thus, the first ``classical'' aspect of the cosmological constant  does not 
seem to be problematic anymore and the classical cosmological constant can have any value, 
being one of the fundamental constants.   

The control of vacuum fluctuations is really important.  
The idea of the (almost) complete cancellation of the vacuum fluctuations 
seems very attractive because it permits to resolve both the cosmological 
and quantum field theoretical problems, connected 
with its treatment. Our idea is very simple. We are inspired by two facts.
\begin{enumerate}
\item the classical equations 
of motion are invariant in respect to the time inversion. 
\item  gravity being 
reparametrisation-invariant theory, does not have a time \cite{rep-grav}. 
Indeed, at least for the closed cosmological models
the Hamiltonian of the theory is equal to zero and the na\"\i ve notion of time looses sense. 
An effective time arises in the process 
of interaction with matter and due to the breakdown of the gauge (reparametrisation) 
invariance due to the gauge fixing choice -- there is ample literature devoted 
to this topic \cite{time-grav}.
\end{enumerate}

Hence, we suggest the following postulate:
the vacuum state evolves time-symmetrically according to the evolution operator

\begin{equation}
W(t) = \frac{1}{2}\,\Big(T\,e^{-iHt}\,T^{-1}\,e^{-iHt}\ +\ e^{-iHt}\,T\,e^{-iHt}\,T^{-1}\Big)
\label{postulate} 
\end{equation}

where $H$ is the Hamiltonian and $T$ is the operator of time inversion.
If the theory is invariant with respect to the time inversion operation, {\it i.e.}
 \begin{equation}
T H T^{-1} = H,
\label{postulate1}
\end{equation}
then $W(t) = I$, which corresponds effectively to zero energy of the vacuum state.   
 
In some respect this situation can be described in terms of negative (mirror) matter - 
some kind of shadow matter. 
The presence of two replicas of fields, having opposite signs of the vacuum energy 
results in the complete cancellation of vacuum energy in the same sense as the mirror 
energy reflection of ref.~\cite{Sundrum} (see also \cite{elze}).
Much before a similar idea was elaborated by Linde \cite{Linde} --
see also \cite{Henry,Moffat}, where the idea of the second negative energy world was put forward.
Certain hints from Superstring theory for shadow  matter with negative vacuum energy were established in \cite{tseytlin}.
Thus, its contribution to the effective cosmological constant vanishes.

Therefore our prescription (\ref{postulate}) is equivalent to subtraction of the ground state 
energy only if time reversal holds 
(\ref{postulate1}). In  this it does not coincide with earlier proposals 
\cite{Linde,Sundrum,Henry,Moffat}.

All written above assumes the exact time invariance of the fundamental physical theory. 
However, one could invoke a suitable 
small breaking of the time symmetry. Indeed, the violation of the CP invariance is an 
experimental fact, and the conservation 
of the CPT symmetry implies unavoidably the breakdown of the time symmetry. 
Such type of breakdown could occur even 
spontaneously as was suggested by Tsung Dao Lee in his seminal paper \cite{Lee}. 
So, we are lead to suspect the existence of a 
connection between a small T (or CP) symmetry violation and a small observable 
value of the cosmological constant. 

In a way our approach reminds that of the mirror world or mirror particles -- 
see \cite{Okun} and references theirein. 
The mirror symmetry  is as well known as the symmetry with respect to 
spatial reflections or parity P symmetry. However, the absence of 
significant interactions between mirror particles and normal ones is 
imposed by the phenomenology and not by general principles 
like in the case of T reflection.  

The problems arising in application of the spontaneous T symmetry breaking 
to cosmology and ways of their solution were considered in \cite{Dolgov}.  

It is important to emphasize, that if the connection between the T violation 
and the cosmological constant value does indeed exist, 
then it could be not connected with the presence of the standard CP-breaking 
terms in the CKM matrix, since 
we are interested in the vacuum expectation energy diagrams, to which those terms 
do not give a contribution.  
It would be rather connected with more subtle scheme, which could explain the small 
scale of the observable cosmological 
constant.

The idea of the presence of fields evolving backward in time and co-existing with 
``normal'' fields evolving 
forward in time was used in many different contexts. First of all, one should cite the works 
by Wheeler and Feynman on time symmetric 
electrodynamics \cite{WF} together with the so called transactional interpretation of quantum 
mechanics \cite{Cramer}.
We should emphasize once again that there are no particles moving backward in time 
in our forward-in-time-world. The only influence 
which this time reversed world makes on us is just the presence of vacuum energy in 
the right-hand-side of the Einstein equations.
Moreover, the appearances of such known observable quantum fluctuation effects like 
the Casimir effect could not be influenced 
by the energy reversed world as well because their observability is based on their 
interaction with normal particles, which 
provides boundary conditions responsible for these effects.  

Such an interaction breaks the time symmetric evolution (\ref{postulate}) 
which as we have suggested is valid only for vacuum 
state.

Notice that the idea that the direction of time  can be connected with the existence 
of a cosmological term 
was first put forward by M.P. Bronstein in the context of Friedmann cosmology \cite{Bronstein}.

\section*{Acknowledgement}
We are grateful to A.A. Starobinsky and G. Venturi for fruitful discussions.
The work of A.A. was supported by  Grants SAB2005-0140; RFBR 05-02-17477 
and Programs RNP 2.1.1.1112; LSS-5538.2006.2. A.K. was partially supported by RFBR 
05-02-17450 and  LSS-1157.2006.2.

\end{document}